\documentclass{article}
\usepackage{spconf,amsmath,graphicx}
\usepackage{multirow,booktabs}
\usepackage{hhline}
\usepackage{mathtools}
\usepackage{amsmath,amssymb,amsfonts}
\usepackage{algorithmic}
\usepackage{comment}
\usepackage{subfig}
\usepackage{caption}
\usepackage{graphicx}

\title{GAN-based Super-Resolution and Segmentation of Retinal Layers in Optical coherence tomography Scans}
\name{\normalsize Paria Jeihouni \textsuperscript{b,c}, Omid Dehzangi \textsuperscript{a,b,c}, Annahita Amireskandari \textsuperscript{d}, Ali Rezai \textsuperscript{a,b}, Nasser M. Nasrabadi \textsuperscript{c}} 

\address{\footnotesize\textsuperscript{a}Department of Neuroscience, \textsuperscript{b}Rockefeller Neuroscience Institute, \footnotesize\textsuperscript{c}Lane Department of Computer Science and Engineering, \footnotesize\textsuperscript{d}Ophthalmology and Visual Sciences,~\\\footnotesize West Virginia University, USA
}

\begin{document}
%
\maketitle
\begin{abstract}
In this paper, we design a Generative Adversarial Network (GAN)-based solution for super-resolution and segmentation of optical coherence tomography (OCT) scans of the retinal layers. OCT has been identified as a non-invasive and inexpensive modality of imaging to discover potential biomarkers for the diagnosis and progress determination of neurodegenerative diseases, such as Alzheimer's Disease (AD). Current hypotheses presume the thickness of the retinal layers, which are analyzable within OCT scans, can be effective biomarkers. As a logical first step, this work concentrates on the challenging task of retinal layer segmentation and also super-resolution for higher clarity and accuracy. We propose a GAN-based segmentation model and evaluate incorporating popular networks, namely, U-Net and ResNet, in the GAN architecture with additional blocks of transposed convolution and sub-pixel convolution for the task of upscaling OCT images from low to high resolution by a factor of four. We also incorporate the Dice loss as an additional reconstruction loss term to improve the performance of this joint optimization task. Our best model configuration empirically achieved the Dice coefficient of 0.867 and mIOU of 0.765.\end{abstract}
\section{Introduction}
\label{sec:intro}
Optical Coherence Tomography (OCT) is a non-invasive opto-medical diagnostic modality that enabled cross-sectional visualization of the internal structure of biological components \cite{schmitt1999optical}, prominently the human retina \cite{fercher2003optical}. According to several clinical studies, the neurodegenerative processes, such as in Alzheimer's, propelled by the abnormal cerebral accumulation of Amyloid-beta and tau protein \cite{ferreira2011neuroimaging}, also may affect the retina. These studies hypothesize the neuronal loss of retinal tissue as a possible biomarker for the presence of AD (i.e. retina layers thicknesses) \cite{bayhan2015evaluation}. Currently, measures such as Position Emission Tomography and Magnetic Resonance Imaging are the standards for AD diagnosis \cite{ferreira2011neuroimaging}. Though, they come with the added burden of being invasive. Thus there is ongoing research regarding the viability of OCT scans as an alternative, as it offers the benefit of being non-invasive, less time consuming and cost-effective as well.  For further research in addressing this modality as a viable biomarker, the challenging task of the retinal layer segmentation is the first significant step. Besides the presence of micro-saccading eye movements, another hindrance commonly faced is the near non-visibility of the layer boundaries, which compels the research of super resolving the images for improved clarity.

Since the advent of neural networks as proven methods for effective application in computer vision tasks \cite{krizhevsky2012imagenet}, there have been numerous developments for the target of semantic segmentation of biomedical images. Most semantic segmentation algorithms follow an encoder-decoder based architecture, popularized by the work \cite{long2015fully} called Fully Convolutional Network (FCN). 
An issue faced with FCN is that successive downsampling and upsampling result in losing some semantic and spatial information. U-Net \cite{ronneberger2015u} solved that issue by introducing skip connections in between the encoder and decoders, to relay the spatial information from the encoder part to the corresponding feature maps of the decoder region.

 U-Net has been widely used in the domain of biomedical image segmentation and has spawned several variations, such as U-Net++ \& 3D-U-Net \cite{zhou2018unet++,cciccek20163d,3rd,4th}. In the specific OCT segmentation related task, ReLayNet \cite{roy2017relaynet} was published which follows the U-Net baseline and to the best of our knowledge, provides the state of the art performance. In this work, we compared our best-achieved results with the ReLayNet method.

GANs have been quite prominent in learning deep representations and modelling high dimensional data. With the advent of conditional GANs \cite{goodfellow2014generative}, it became possible to capture even better representations, by rendering both the generator and discriminator networks as class conditionals, and showed good performance translating data from one domain to another \cite{isola2017image, zhu2017unpaired}, thus being appropriate for semantic segmentation. Super resolution is another long challenging task in the computer vision domain, which aims in constructing high-resolution photo realistic images from their low-resolution counterparts. In Super Resolution Convolutional Neural Network (SR-CNN) \cite{dong2015image}, the image is first upsampled through bi-cubic interpolation, and fed through an FCN, resulting in output with high resolution. The work in \cite{ledig2017photo}  is the continuation of SR-CNN, with residual blocks replacing the conventional convolution blocks. Using such SR-CNN as the generator architecture, GANs have been also used to reconstruct images in higher resolution \cite{ledig2017photo}.



In this work, we have identified the goal as jointly superresolving and segmenting the OCT retinal scans. We design a conditional GAN with different generator architectures as well as analyzing the effect of a Dice loss as an additional constraint, and how its presence improves the performance.

\begin{figure}[t]
\centering
\includegraphics[width=8.6cm]{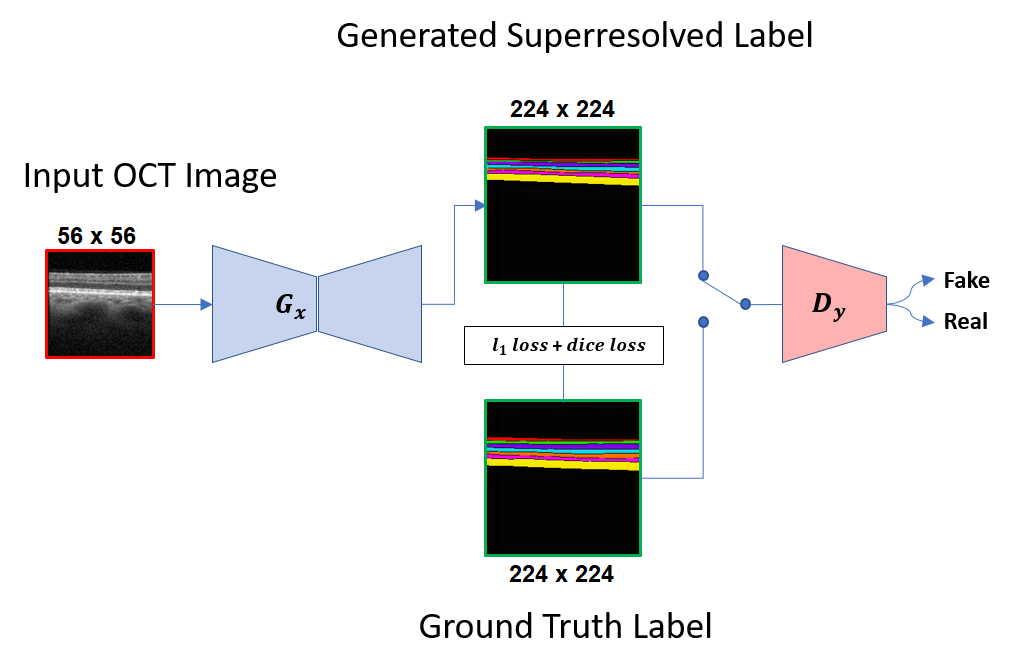}
\caption{The high level GAN architecture for super-resolution and segmentation, where input images of size $56 \times 56$ are fed to the generator $G_{x}$ that outputs realistic superesolved segmented images of $224 \times 224$ to bluff the discriminator, $D_{y}$.}
\label{ganArc}
\end{figure}

\section{Data Acquisition and pre-processing}
For experimentation, OCT images of 45 patients were obtained from the Department of Opthalmology \& Visual Sciences, West Virginia University. Nineteen scans were captured from each patient utilizing the Spectralis OCT imaging platform by Infinity. For this research work, seven layers of the retina were focused on which are Internal Limiting Membrane (ILM), RNFL, GCL, Inner Plexiform Layer (IPL), Inner Nuclear Layer (INL), Outer Plexiform Layer (OPL), and Outer Nuclear Layer (ONL). The OCT images which totaled 855, were manually annotated (the 7 layers and the background) by an expert in this domain.

We applied data augmentation techniques to synthetically enlarge the dataset. The techniques being horizontal flip, rotation (15 degrees), and spatial translation. Apart from these conventional augmentation methods, the dataset was subjected to a sliding crop window with 75\% overlap at each sliding step, effectively increasing the dataset by a significant factor. Similar augmentations were also done to the ground truth labels. After cropping, each patch were of the size of $224 \times 224$. The presence of speckle noise is a big hindrance as it corrupts the edges between the retinal layers. To alleviate this issue, a median filter of a 3x3  window was used. On top of that, an unsharp masking technique is used to make the boundaries more visible for the task at hand. 

To get the labels for the tasks of segmentation and surer-resolution at the same time, we referred to the accurate labels of size (224x224) as ground-truth target labels and down-sampled OCT scans to the size of 56x56 as the input. The input $56 \times 56$ images were fed to the generator $G$ that generates the segmented outputs of size $224 \times 224$, upscaled by a factor of four, to compare against the target labels. 
\begin{figure}[!t]
\centering
\subfloat[Bottleneck Residual Block]{\includegraphics[width=0.12\textwidth]{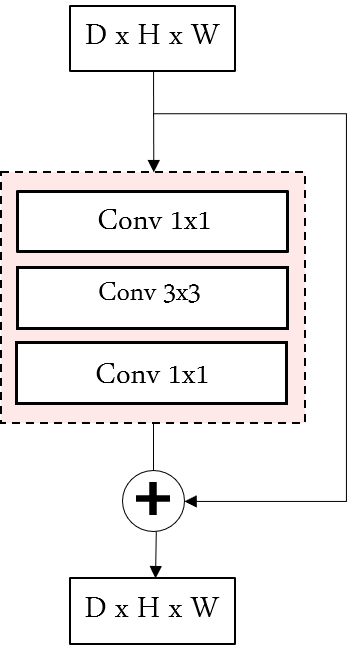}%
\label{fig_first_res}}
\hfil
\subfloat[Transposed Bottleneck Residual Block]{\includegraphics[width=0.2\textwidth]{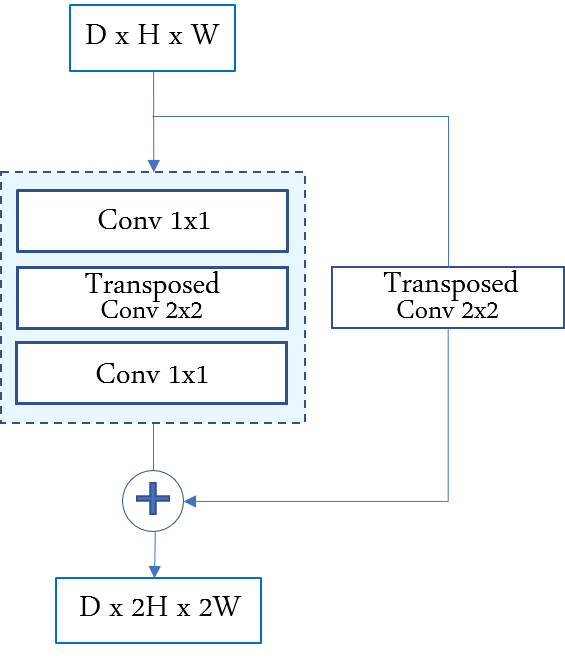}%
\label{fig_second_res}}
\caption{Two types of Residual Blocks. (a) a typical bottleneck block from \cite{he2016deep}. The $1 \times 1$ convolutions decrease and increase the depth of the feature maps to lead to more efficient computation. (b) the bottleneck block with transposed convolution block in the middle instead of a typical convolution block. }
\label{fig_resblock}
\end{figure}
\section{Methodology}
The baseline architecture of a GAN consists of two competing networks, aptly named generator and the discriminator. The purpose of the generator in this work is to produce superresolved segmented labels of the OCT input images, whereas the discriminator learns to differentiate between real ground truth labels and the generated ones. Fig \ref{ganArc} shows the high-level architecture of the GAN. The following chronicles each component of the architecture in detail.

\subsection{Generator}
For our designed GAN, we employed two different architectures, namely, U-net and ResNet, with two different upsampling modules, namely, transposed and sub-pixel convolutions to undertake the dual task of segmenting the input OCT images and superresolving them.


\subsubsection{ResNet}
ResNet architecture was one of the most groundbreaking works in computer vision \cite{he2016deep}, which introduced the significant concept of skip connections. In Fig \ref{fig_first_res}, an example residual block is shown where a single $3 \times 3$ convolution is stacked between two $1 \times 1$ convolutions, with the input to the block is bypassed and added to its output. Connecting these blocks to one another, ResNet of varied sizes is formed. 

\begin{figure}[!t]
\centering
\subfloat[ResNet with transposed convolution]{\includegraphics[width=0.2\textwidth]{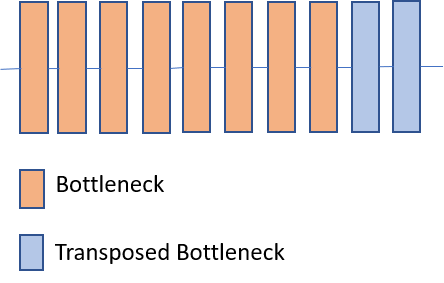}%
\label{fig_first_resT}}
\hfil
\subfloat[ResNet with sub-pixel convolution]{\includegraphics[width=0.2\textwidth]{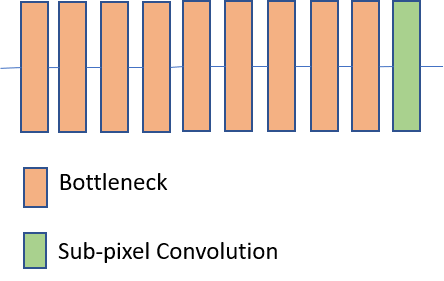}%
\label{fig_second_resT}}
\caption{The ResNet architectures with a series of residual bottlenecks that, at the end, upscales the feature map using: (a) two transposed bottlenecks , each by a factor of 2, and (b) a serial sub-pixel convolution connection layer (factor of 4).}
\label{resGen}
\end{figure}

For our task, we evaluated two block types to upsample the low-resolution feature maps to high resolution: 1) To keep in line with the residual blocks, two modules of transposed block depicted in Fig \ref{fig_second_res} is used for the upsizing operation. in between the 1x1 convolutions, a 2x2 transposed convolution is used which acts as 2-fold upsampler. The residual connection also goes through a transposed convolution to maintain spatial integrity. 2) We also evaluated sub-pixel convolution popularized by \cite{shi2016real}. An image with $Cr^2\times H\times W$ dimensions where $C$ is the channel width, $H$ and $W$ being the spatial height and width, the sub pixel convolution of it will yield an output of the size $C \times Hr \times Wr$, where $r$ is the factor by which the image is being upscaled. We add a sub-pixel convolution block, at the end, with an upscale factor $r$ of 4 (both blocks shown in Fig. \ref{resGen}).
\subsubsection{U-Net}
\begin{figure}[!t]
\centering
\subfloat[With Transposed Convolution]{\includegraphics[width=0.42\textwidth]{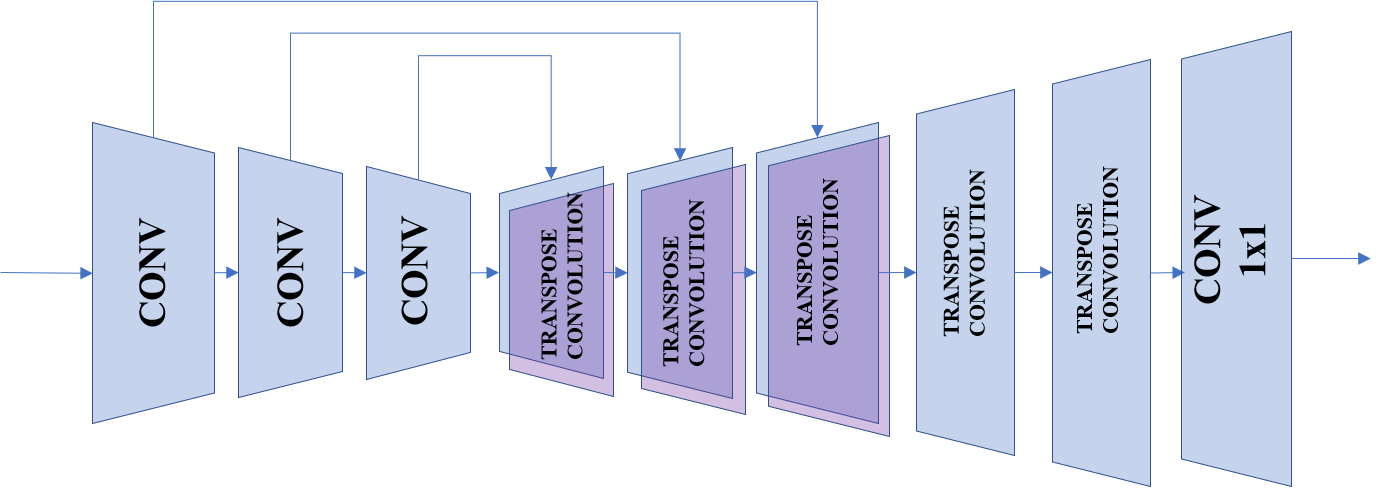}%
\label{fig_first_case}}
\hfil
\subfloat[With sub-pixel Convolution]{\includegraphics[width=0.42\textwidth]{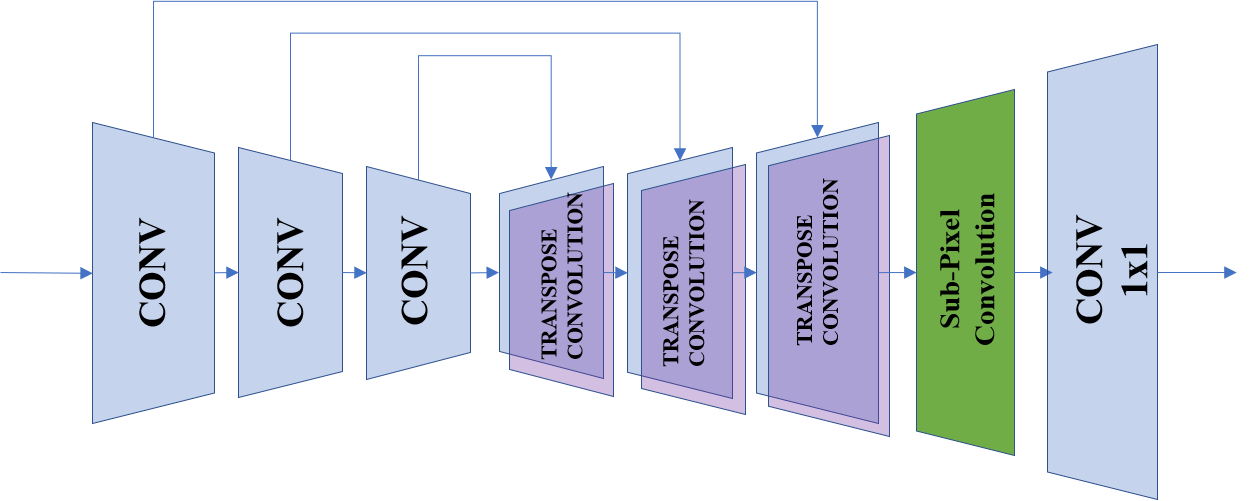}%
\label{fig_second_case}}
\caption{U-Net architectures with down/upsampling layers and the the skip connections in between. Then, two additional upsampling techniques: (a) two layers of transposed convolution, and (b) a sub-pixel convolution. At the end, a $1 \times 1$ convolution layer reduces the feature map depth to 3 (for RGB).}
\label{unetArc}
\end{figure}
U-Net follows a common encoder-decoder architecture with the presence of skip connections to bridge the encoder and decoder portions. That results in retrieving feature representations extracted during the encoding operation. 
For the task of super-resolution, we compared transposed vs. sub-pixel convolution blocks. The former has $2 \times 2$ kernels with the stride of 2, so that each block upscales by a factor of 2. Also, the aforementioned former block was used for upsampling with a factor $r$ as 4 (shown in Fig \ref{unetArc}a \& \ref{unetArc}b).

\subsection{Discriminator}
To differentiate real high-resolution images from the generated ones, we use a patchGAN classifier \cite{isola2017image}. The network consists of several blocks, each including a convolution layer, a relu activation function,  and a batch normalization layer, which successively decrease the spatial size of the input to $N \times N$ size patches to classify the said input as actual ground truth labels or the generated ones. For our experiment, the patch size was chosen as $70 \times 70$. 

\subsection{Loss Function}
In this work, we have opted three different loss functions for the purpose of training the algorithm, including the adversarial loss, the generator reconstruction loss, and also the addition of the Dice loss. The combination of all three contributes to the backpropagation and update of the model weights. The adversarial loss is applied on both the generator $G$ and the discriminator $D$. If the input to the $G$ is $x$, then the adversarial loss is
\begin{equation}
    L_{GAN}(G) = E_{x}[log(D(G(x)))],
\end{equation}
and for training the discriminator, the loss function being:
\begin{equation}
    L_{GAN}(G,D) = E_{y}[logD(y)] + E_{x}[log(1-D(G(x))],
\end{equation}
where the $y$ denotes the actual ground truth label. 
\noindent The reconstruction loss is the L1 loss measured between the generated output $G(X)$ and the ground truth label $Y$ and is given as
\begin{equation}
   L_{L1}(G) = E_{x}[\|y-G(x)\|]_{1}.
\end{equation}
This loss function helps the generator in synthesizing output conforming to the ground truth label, appropriately training the generator for the desired task.
\subsubsection{Dice Loss}
This loss function originates from the semantic segmentation metric called the Dice coefficient. Taking the additive inverse of said metric gives us the Dice loss as in  \cite{milletari2016v}. The task of the network is to minimize this function so that the generator can successfully segment the images which results in minimal Dice loss when calculated against the ground truths. This loss function acts as an additional reconstruction loss to further emphasize and improve the quality of the generator output.
The total loss function for the generator stands as
\begin{equation}
    L_{generator} = L_{GAN}(G) + \lambda L_{L1}(G) + \alpha L_{Dice}(G),
\end{equation}
where $ \lambda $ and $ \alpha $ are coefficients of constant value, which control the relative importance of each corresponding loss function that we fine-tuned by a grid-search that maximizes the Dice and mIOU values.

\section{Experiments and Results}
Experiments are conducted according to the baseline architecture in Fig. \ref{ganArc}. After the preprocessing and augmentation, the dataset size was increased from 855 to 34,199. The dataset was split into training and test set, with the training set containing 80\% of the total data. All of the training was done on a system with two GeForce GTX TITAN X GPU. Adam optimizer \cite{kingma2014adam} was used for training, with a learning rate of 0.0001 for both the generator and the discriminator. The $ \lambda $ value was set to 100 and $ \alpha $ was set to 1. We first compared the results of U-net and ResNet to produce segmentation in the $56 \times 56$ resolution with a) no super-resolution (SR) and b) with SR-CNN to obtain the high-resolution version, dis-jointly. Then, we evaluated the performance gain of our proposed model over the above. Hence, the four architectures of U-net and ResNet, each with transposed vs. subpixel conv. were evaluated and compared as our proposed generator. For all cases, we also evaluated the effect of Dice loss as an additional cost term. All of the experiments ran for 100 epochs.

The metric used for the quality of the superresolved segmentation was the Dice coefficient and mean intersection over union (mIOU), which are of the $[0,1]$ range where a higher value denotes better quality. The Dice coefficient was chosen over pixel accuracy because the latter does not take into account the problem of class imbalance (dominant background).


\begin{table}[]
\centering
\caption{Segmentation performance for various generator, $G$, architectures with no SR vs. a disjoint SR-CNN,  with/without the addition of the Dice loss as an objective term.}


\label{tab:table-1}
\resizebox{0.5\textwidth}{!}{%
\begin{tabular}{|c|c|c|c|}
\hline
\multicolumn{2}{|c|}{\textbf{Model} }               & \textbf{Dice Coefficient} & \textbf{mIOU} ~\\ \hline
No Dice & GAN(U-Net)       & 0.816                     & 0.685         ~\\ \hline
No Dice & GAN (U-Net) then SR-CNN              & 0.814                     & 0.681         ~\\ \hline
No Dice & GAN (ResNet) & 0.825 & 0.690         ~\\ \hline
No Dice & GAN (ResNet) then SR-CNN   & 0.831 & 0.692        ~\\ \hhline{|=|=|=|=|}
Added Dice &  GAN (U-Net)               & 0.822                     & 0.710         ~\\ \hline
Added Dice & GAN (U-Net) then SR-CNN              & 0.825                     & 0.709         ~\\ \hline
Added Dice & GAN (ResNet)   & 0.833                     & 0.718         ~\\ \hline
\textbf{Added Dice} & \textbf{GAN (ResNet) then SR-CNN}  & \textbf{0.838}                     & \textbf{0.721}         ~\\ \hline
\end{tabular}%
}
\end{table}

\begin{table}[]
\centering
\caption{Performance comparison for different $G$ architectures with joint SR and segmentation, w/wo the Dice loss term.}
\label{tab:table-2}
\resizebox{0.5\textwidth}{!}{%
\begin{tabular}{|c|c|c|c|}
\hline
\multicolumn{2}{|c|}{\textbf{Model} }               & \textbf{Dice Coefficient} & \textbf{mIOU} ~\\ \hline
No Dice & GAN(U-Net + Transposed\_Conv)       & 0.834                     & 0.719         ~\\ \hline
No Dice & GAN(U-Net + Sub-pixel\_Conv)              & 0.831                     & 0.718         ~\\ \hline
No Dice & GAN(ResNet + Transposed\_Conv) & 0.840 & 0.729         ~\\ \hline
No Dice & GAN(ResNet + Sub-pixel\_Conv)   & 0.855 & 0.743        ~\\ \hhline{|=|=|=|=|}
Added Dice &  GAN(U-Net + Transposed\_Conv)               & 0.839                     & 0.722         ~\\ \hline
Added Dice & GAN(U-Net + Sub-pixel\_Conv)               & 0.841                     & 0.734         ~\\ \hline
Added Dice & GAN(ResNet + Transposed\_Conv) & 0.853                     & 0.745         ~\\ \hline
\textbf{Added Dice} & \textbf{GAN(ResNet + Sub-pixel\_Conv)}   & \textbf{0.867}                     & \textbf{0.765}         ~\\ \hhline{|=|=|=|=|}
Added Dice & ReLayNet \cite{roy2017relaynet}  & 0.856                     & 0.751        ~\\ \hline
\end{tabular}%
}
\end{table}

\subsection{Comparative results}
Table \ref{tab:table-1} reports the comparative results between U-net and ResNet with no SR, a disjoint SR-CNN, and Table \ref{tab:table-2} shows the results achieved by our proposed joint SR and segmentation (w/wo the Dice term). Table \ref{tab:table-1} demonstrate a minor improvement that is not consistent either by adding the disjoint pre-trained SR-CNN module that further motivates the need for jointly optimizing the two tasks. Table \ref{tab:table-1} and Table \ref{tab:table-2} report consistent enhanced results of ResNet over U-net, sub-pixel over transpose conv, and consistent improvement of the added Dice term over no Dice overall. It is worth mentioning that training of ResNet as $G$ took longer than than U-net  as $G$. Comparing Table \ref{tab:table-1} and Table \ref{tab:table-2}, it is evident that a superior performance is observed by our proposed model over disjoint SR, by optimizing SR and segmentation in one whole architecture. Given our dataset, Table \ref{tab:table-2} also reports that our best model configuration, "AddedDice+GAN(ResNet+sub-pixel\_Conv)", generates $7.64\%$ relative improvement over the Dice coefficient achieved by the ReLayNet as a state of the art OCT segmentation model.
\section{Conclusion}
Our aim in this paper was to generate superresolved segmentation for OCT scans of the retina using our proposed GAN architectures. We experimented with various architectures as generators that performed the dual task of semantic segmentation as well as superresolving the segmented images. To do this dual training, we deployed two popular architectures, U-Net and ResNet, with additional blocks of transposed convolution vs. sub-pixel convolution for the task of upscaling images from low to high resolution. We also investigated incorporating the Dice loss, an objective function originating from the Dice coefficient metric, as an additional loss function for the GAN model. As evident from the results, the joint training for the dual task of segmentation and super-resolution provided effective result enhancement. The inclusion of the Dice loss emphasized the reconstruction performance and improved the empirical results consistently. 

\bibliographystyle{IEEEbib}
\bibliography{report.bib}


\end{document}